\documentclass[conference, 10pt, final]{IEEEtran}

\usepackage{enumitem}
\usepackage[nolist]{acronym} 
\usepackage{tikz}
\usepackage{pgfplots}
\usepgfplotslibrary{groupplots}
\usepackage{graphicx}
\usepackage{hyperref}
\usepackage{subcaption}
\usepackage{float}
\pgfplotsset{compat=newest}
\usepackage{units}
\usepackage{amsmath, amsbsy, amssymb, amsthm}
\usepackage{tikzscale}
\usetikzlibrary{arrows}
\usepackage{color}
\usepackage{epigraph}
\usepackage{circuitikz}
\usepackage{multirow}
\usepackage{booktabs}
\usepackage[plainruled]{algorithm2e}
\usepackage[group-separator={,}]{siunitx}
\definecolor{darkgreen}{rgb}{0.125,0.5,0.169}
\usetikzlibrary{shapes,arrows}
\usetikzlibrary{positioning}
\usetikzlibrary{patterns}
\usetikzlibrary{decorations.pathreplacing}
\usepackage{marvosym}
\usepackage{bbm}
\usepackage{mathtools}
\usepackage{xfrac}
\usepackage{booktabs}
\usepackage{array}

\tikzset{>=latex}

\begin{acronym}
 \acro{R2M}{raw 2\textsuperscript{nd} moment}
 \acro{CSI}{channel state information}
 \acro{UE}{user equipment}
 \acro{UL}{uplink}
 \acro{BS}{base station}
 \acro{TDD}{time division duplex}
 \acro{FDD}{frequency division duplex}
 \acro{ECC}{error-correcting code}
 \acro{MLD}{maximum likelihood decoding}
 \acro{HDD}{hard decision decoding}
 \acro{IF}{intermediate frequency}
 \acro{RF}{radio frequency}
 \acro{SDD}{soft decision decoding}
 \acro{NND}{neural network decoding}
 \acro{CNN}{convolutional neural network}
 \acro{ML}{maximum likelihood}
 \acro{GPU}{graphical processing unit}
 \acro{BP}{belief propagation}
 \acro{LTE}{Long Term Evolution}
 \acro{BER}{bit error rate}
 \acro{SNR}{signal-to-noise-ratio}
 \acro{ReLU}{rectified linear unit}
 \acro{BPSK}{binary phase shift keying}
 \acro{QPSK}{quadrature phase shift keying}
 \acro{AWGN}{additive white Gaussian noise}
 \acro{MSE}{mean squared error}
 \acro{LLR}{log-likelihood ratio}
 \acro{MAP}{maximum a posteriori}
 \acro{NVE}{normalized validation error}
 \acro{BCE}{binary cross-entropy}
 \acro{CE}{cross-entropy}
 \acro{BLER}{block error rate}
 \acro{SQR}{signal-to-quantisation-noise-ratio}
 \acro{MIMO}{multiple-input multiple-output}
 \acro{OFDM}{orthogonal frequency division multiplex}
 \acro{RF}{radio frequency}
 \acro{LOS}{line of sight}
 \acro{NLoS}{non-line of sight}
 \acro{NMSE}{normalized mean squared error}
 \acro{CFO}{carrier frequency offset}
 \acro{SFO}{sampling frequency offset}
 \acro{IPS}{indoor positioning system}
 \acro{TRIPS}{time-reversal IPS}
 \acro{RSSI}{received signal strength indicator}
 \acro{MIMO}{multiple-input multiple-output}
 \acro{ENoB}{effective number of bits}
 \acro{AGC}{automatic gain control}
 \acro{ADC}{analog to digital converter}
 \acro{ADCs}{analog to digital converters}
 \acro{FB}{front bandpass}
 \acro{FPGA}{field programmable gate array}
 \acro{JSDM}{Joint Spatial Division and Multiplexing}
 \acro{NN}{neural network}
 \acro{IF}{intermediate frequency}
 \acro{LoS}{line-of-sight}
 \acro{NLoS}{non-line-of-sight}
 \acro{DSP}{digital signal processing}
 \acro{AFE}{analog front end}
 \acro{SQNR}{signal-to-quantisation-noise-ratio}
 \acro{SINR}{signal-to-interference-noise-ratio}
 \acro{ENoB}{effective number of bits}
 \acro{PCB}{printed circuit board}
 \acro{EVM}{error vector mangnitude}
 \acro{CDF}{cumulative distribution function}
 \acro{MRC}{maximum ratio combining}
 \acro{MRP}{maximum ratio precoding}
 \acro{MRT}{maximum ratio transmission}
 \acro{DeepL}{deep-learning}
 \acro{DL}{downlink}
 \acro{SISO}{single-input single-output}
 \acro{SGD}{stochastic gradient descent}
 \acro{CP}{cyclic prefix}
 \acro{MISO}{Multiple Input Single Output}
 \acro{LMMSE}{linear minimum mean square error}
 \acro{ZF}{zero forcing}
 \acro{USRP}{universal software radio peripheral}
 \acro{RNN}{recurrent neural network}
 \acro{GRU}{gated recurrent unit}
 \acro{LSTM}{long short-term memory}
 \acro{NTM}{neural turing machine}
 \acro{DNC}{differentiable neural computer}
 \acro{TCN}{temporal convolutional network}
 \acro{FCL}{fully connected layer}
 \acro{MANN}{memory augmented neural network}
 \acro{RNN}{recurrent neural network}
 \acro{DNN}{dense neural network}
 \acro{FIR}{finite impulse response}
 \acro{BPTT}{back-propagation through time}
 \acro{GAN}{generative adversarial network}
 \acro{ELU}{exponential linear unit}
 \acro{tanh}{hyperbolic tangent}
 \acro{BICM}{bit-interleaved coded modulation}
 \acro{OTA}{over-the-air}
 \acro{IM}{intensity modulation}
 \acro{DD}{direct detection}
 \acro{RL}{reinforcement learning}
 \acro{SDR}{software-defined radio}
 \acro{WGAN}{Wasserstein generative adversarial network}
 \acro{BMD}{bit-metric decoding}
 \acro{BMI}{bit-wise mutual information}
 \acro{LDPC}{low-density parity-check}
 \acro{IDD}{iterative demapping and decoding}
 \acro{JSD}{Jensen-Shannon divergence}
 \acro{MMSE}{minimum mean square error}
 \acro{FFT}{fast Fourier transform}
 \acro{IFFT}{inverse fast Fourier transform}
 \acro{QAM}{quadrature amplitude modulation}
 \acro{EMD}{earth mover's distance}
 \acro{TDL}{tapped delay line}
 \acro{KL}{Kullback-Leibler}
 \acro{PRACH}{physical random access channel}
 \acro{URLLC}{ultra-reliable low-latency communication}
 \acro{ANOMA}{asynchronous non-orthogonal multiple access}
 \acro{FEC}{forward error correction}
 \acro{PAPR}{peak-to-average power ratio}
 \acro{APP}{a posteriori probability}
 \acro{COTS}{commercial off-the-shelf}
 \acro{PLL}{phase locked loop}
 \acro{STO}{sampling time offset}
 \acro{SFO}{sampling frequency offset}
 \acro{CFO}{carrier frequency offset}
 \acro{CPO}{carrier phase offset}
 \acro{CSI}{channel state information}
 \acro{GNSS}{global navigation satellite system}
 \acro{ELAA}{extremely large aperture array}
 \acro{UE}{user equipment}
 \acro{DICHASUS}{\underline{Di}stributed \underline{Cha}nnel \underline{S}ounder by \underline{U}niversity of \underline{S}tuttgart}
 \acro{JCAS}{Joint Communication and Sensing}
 \acro{CT}{Continuity}
 \acro{TW}{Trustworthiness}
 \acro{KS}{Kruskal Stress}
 \acro{PCA}{Principal Component Analysis}
\end{acronym}

\definecolor{mittelblau}{RGB}{0, 126, 198}
\definecolor{violettblau}{cmyk}{0.9, 0.6, 0, 0}
\definecolor{rot}{RGB}{238, 28 35}
\definecolor{apfelgruen}{RGB}{140, 198, 62}
\definecolor{gelb}{RGB}{1, 221, 0}
\definecolor{orange}{RGB}{244, 111, 33}
\definecolor{pink}{RGB}{237, 0, 140}
\definecolor{lila}{RGB}{128, 10, 145}
\definecolor{hellgrau}{RGB}{224, 224, 224}
\definecolor{mittelgrau}{RGB}{128, 128, 128}
\definecolor{dunkelgrau}{RGB}{80,80,80}
\definecolor{anthrazit}{RGB}{19, 31, 31}

\definecolor{bgorange}{HTML}{fcc0a7}
\definecolor{bggreen}{HTML}{ccebb9}

\IEEEoverridecommandlockouts

\begin{document}

\title{Improving Triplet-Based Channel Charting on Distributed Massive MIMO
Measurements}

\author{\IEEEauthorblockN{Florian Euchner, Phillip Stephan, Marc Gauger, Sebastian D\"orner, Stephan ten Brink \\}

\IEEEauthorblockA{
Institute of Telecommunications, Pfaffenwaldring 47, University of  Stuttgart, 70569 Stuttgart, Germany \\ \{euchner,stephan,gauger,doerner,tenbrink\}@inue.uni-stuttgart.de
}

}

\maketitle

\begin{abstract}
The objective of channel charting is to learn a virtual map of the radio environment from high-dimensional \ac{CSI} that is acquired by a multi-antenna wireless system.
Since, in static environments, \ac{CSI} is a function of the transmitter location, a mapping from \ac{CSI} to channel chart coordinates can be learned in a self-supervised manner using dimensionality reduction techniques.
The state-of-the-art triplet-based approach is evaluated on multiple datasets measured by a distributed massive \ac{MIMO} channel sounder, with both co-located and distributed antenna setups.
The importance of suitable triplet selection is investigated by comparing results to channel charts learned from a genie-aided triplet generator and learned from triplets on simulated trajectories through measured data.
Finally, the transferability of learned forward charting functions to similar, but different radio environments is explored.
\end{abstract}

\acresetall

\section{Introduction}
Motivated by ever increasing wireless traffic volumes, spatial multiplexing through massive \ac{MIMO} has been identified as a crucial technology for improving spectral efficiency.
With massive \ac{MIMO}, which requires a large number of antennas at the \ac{BS}, the number of channel estimates necessary for communication over the channel between \ac{BS} and \ac{UE} has increased.
Acquired channel estimates, which can be expressed in various representations (e.g., in frequency domain or in time domain), are commonly referred to as \ac{CSI}.
Collecting \ac{CSI}, which necessarily arises at the \ac{BS}, opens the door for many data-driven applications, including localization of \acp{UE}.
Successful supervised learning experiments have established that localization based on \ac{CSI} fingerprinting is possible in principle \cite{savic2015fingerprinting} \cite{vieira2017deep} \cite{Arnold2018} \cite{cc_features} \cite{Arnold2018OnDL}.
However, any supervised learning technique requires accurate \ac{UE} position labels for training, which are usually not available.
Self-supervised training methods, which do not require ground truth position labels, are therefore an attractive alternative.

Channel charting, originally proposed by C. Studer et al. \cite{studer_cc}, is one such self-supervised technique that aims to learn a mapping from the high-dimensional space of possible \ac{CSI} vectors to a low-dimensional space, the so-called channel chart.
The channel chart is supposed to maintain the local geometry of the radio environment.
Depending on the application, it may be desirable to be able to map the chart back to physical space, or the channel chart may be useful in and of itself.

\begin{figure}
    \centering
    \includegraphics[width=0.9\columnwidth]{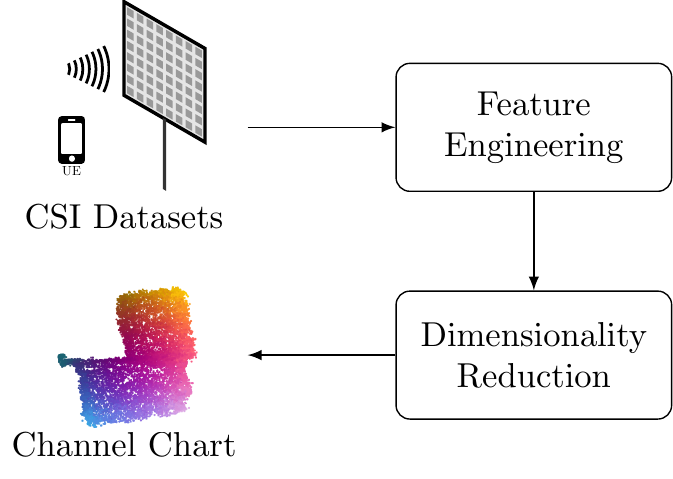}
    \caption{Overview of the Channel Charting Pipeline}
    \label{fig:elements}
\end{figure}

As shown in Fig. \ref{fig:elements}, the key steps in channel charting, apart from acquiring a large \ac{CSI} dataset, are feature engineering and dimensionality reduction.
In a practical system, large \ac{CSI} datasets could be obtained at the \ac{BS} from multiple \acp{UE} over a long timespan.
Despite \ac{CSI} being abundant in real-world massive \ac{MIMO} systems, most of the research currently published is based on synthetic data.
Trials on real-world measurements, on the other hand, are rare and perform poorly in comparison \cite{split_triplet_cc}.
This raises the question of whether this is due to the available amount of data, due to properties of real-world datasets or a result of shortcomings of the current feature engineering or dimensionality reduction methods.
We aim to help address this question by making the following contributions:
\begin{itemize}
    \item With our channel sounder, we measure large \ac{CSI} datasets, which are made publicly available\footnote{Datasets and a tutorial for a special case of channel charting is available at \url{https://dichasus.inue.uni-stuttgart.de/tutorials/tutorial/channelcharting/}}, and apply state-of-the-art triplet neural network-based channel charting, as defined in Section \ref{sec:stateoftheart}, to the data.
    \item We introduce genie-aided and partially genie-aided baselines for triplet selection to benchmark time-based triplet selection against and compare the performance of these different selection rules for \ac{DNN} training in Section \ref{sec:tripletgeneration}.
    \item We test the transferability of learned channel mappings to different datasets for same and similar, but different environments in Section \ref{sec:transfer}.
\end{itemize}


\section{State-of-the-Art Channel Charting Overview}
\label{sec:stateoftheart}

\subsection{System Model}
We consider a wireless transmission system consisting of a massive \ac{MIMO} \ac{BS} with $B$ antennas and a single-antenna transmitter.
At the \ac{BS}, \ac{CSI} for $W$ \ac{OFDM} subcarriers is collected at each time instant $n = 1, \ldots, N$ for different \ac{UE} positions $\mathbf{x}_n \in \mathbb{R}^D$, with $D$ being the physical spatial dimensionality.
\ac{CSI} for a particular time instant can be expressed either as a matrix of channel coefficients $\mathbf S_n \in \mathbb C^{B \times W}$, or in its vectorized representation $\mathbf h_n = \mathrm{vec}(\mathbf S_n) \in \mathbb C^M$, with $M = B \cdot W$.
As part of a feature engineering stage $\mathcal F: \mathbb C^M \to \mathbb C^{M'}$, the \ac{CSI} vector is transformed into a feature vector $\mathbf{f}_n \in \mathbb{R}^{M'}$.

Channel charting is a dimensionality reduction technique, with the objective of finding a \emph{forward charting function}
\[
    \mathcal C: \mathbb{C}^{M'} \rightarrow \mathbb{R}^{D'},
\]
which maps points from feature space to the $D'$-dimensional channel chart (here: $D' = 2$).
Charts are generated from a dataset made up of datapoints that are 3-tuples of channel coefficients $\mathbf{S}_{n}$, \ac{UE} positions $\mathbf x_n \in \mathbb R^D$ and timestamps $t_n$:
\[
    \text{Dataset:} ~ \left\{ (\mathbf{S}_{n}, \mathbf x_n, t_n) \right\}_{n = 1, \ldots, N}
\]


\subsection{Feature Engineering}
\label{sec:featureengineering}
The main purpose of feature engineering is to extract large-scale fading characteristics from \ac{CSI}.
In this work, we adopt the \emph{scaled \ac{R2M}} representation $\tilde {\mathbf H} \in \mathbb C^{M \times M}$ as originally defined in \cite{cc_features} (including their notation), and choose the estimated path loss exponent to be $\sigma = 8$.
Here, the feature vector $\mathbf f_{n, \mathrm{cplx}}$ is the vectorized form of $\tilde {\mathbf H}$, i.e. $\mathbf f_{n, \mathrm{cplx}} = \mathrm{vec}(\tilde {\mathbf H}) \in \mathbb C^{M'}$ and $M' = M^2$.
Experiments on our datasets suggest that using either only real or only imaginary parts of $\mathbf f_{n, \mathrm{cplx}}$ yields a performance comparable to using complex-valued \ac{R2M}.
Therefore, and for complexity reasons, we chose the final feature vector to be $\mathbf{f}_n = \Re\{\mathbf f_{n, \mathrm{cplx}}\}$.

\subsection{Forward Charting Function}
The forward charting function $\mathcal C$ can either be implemented as a conventional dimensionality reduction technique or as a trainable \ac{DNN}.
By realizing $\mathcal C$ as a \ac{DNN}, once trained, new \ac{CSI} datapoints can be mapped to the channel chart with low computational complexity, which is highly desirable.
The \ac{DNN} may be trained using an autoencoder structure \cite{studer_cc}, as part of a siamese network \cite{siamese_cc} or using triplet loss \cite{triplet_cc}.
We will focus on the last, which appears to be most promising.

Furthermore, charting functions can be learned either \emph{purely} based on \ac{CSI}, or based on \ac{CSI} and timestamps:
Without relying on ground truth position labels, it is difficult to tell whether two \ac{CSI} samples are close to each other in physical space or not.
However, based on the assumption that samples measured close in time are likely to be close in space as well, \cite{triplet_cc} proposes to select triplets based on timestamps, which are almost certainly available at any \ac{BS}.
Since the availability of timestamp labels does not pose a challenge for practical systems, we will focus on training techniques which do take time into account.

\subsection{Triplet loss-based \ac{DNN} training}
Figuratively speaking, a charting function $\mathcal C$ is good if it preserves the local and global geometry of real space except for rotations and/or scalings:
Datapoints that are close to each other in physical space should also be close in the channel chart (and vice versa for distant datapoints).
This motivates training with triplet loss, where the \ac{DNN} $\mathcal C$ learns from \emph{positive} (similar) and \emph{negative} (dissimilar) examples \cite{triplet_cc}.
For training, three \ac{CSI} vectors measured at physical locations $\left(\mathbf{x}_\mathrm{anchor},\mathbf{x}_\mathrm{pos},\mathbf{x}_\mathrm{neg}\right)$ are required.
This triplet consists of the anchor point $\mathbf{x}_\mathrm{anchor}$, the positive sample $\mathbf{x}_\mathrm{pos}$ and the negative sample $\mathbf{x}_\mathrm{neg}$.
For the subsequent investigations, it is important to know that $\mathbf{x}_\mathrm{anchor}$, $\mathbf{x}_\mathrm{pos}$ and $\mathbf{x}_\mathrm{neg}$ should fulfill
\begin{equation}
    \lVert \mathbf{x}_\mathrm{anchor} - \mathbf{x}_\mathrm{pos} \rVert \leq \lVert \mathbf{x}_\mathrm{anchor} - \mathbf{x}_\mathrm{neg} \rVert.
    \label{eq:distance_inequality}
\end{equation}
Finding such triplets is the subject of \emph{triplet selection}.


\subsubsection{Triplet Selection}
To generate a set of --- in our case $1\,200\,000$ --- triplets to train the \ac{DNN} on, an anchor datapoint $(\mathbf S_\mathrm{anchor}, \mathbf x_\mathrm{anchor}, t_\mathrm{anchor})$ is drawn from the dataset for each triplet.
Then, a datapoint $(\mathbf S_\mathrm{pos}, \mathbf x_\mathrm{pos}, t_\mathrm{pos})$ is randomly chosen out of the set of datapoints $(\mathbf S_n, \mathbf x_n, t_n)$ for which $|t_n - t_\mathrm{anchor}| \leq T_\mathrm{c}$, where $T_\mathrm{c}$ is a threshold interval for positive samples.
Unless otherwise specified, we will assume $T_\mathrm{c} = 1.5\,\mathrm{s}$.
Next, the negative sample $(\mathbf S_\mathrm{neg}, \mathbf x_\mathrm{neg}, t_\mathrm{neg})$ is randomly drawn from the entire dataset.
We found that imposing a restriction on the negative sample such as $|t_\mathrm{neg} - t_\mathrm{anchor}| > T_\mathrm{c}$ does not lead to better channel charts, most likely since it is probable that the restriction is fulfilled anyway.
All random selections are performed with uniform probability.
Note that Eq. (\ref{eq:distance_inequality}) is likely to be fulfilled by this triplet choice, but it is not guaranteed to be.

\subsubsection{Neural Network Training}
Our neural network architecture and loss function are identical to \cite[Section VI.B]{triplet_cc}.
The \ac{DNN} $\mathcal C$ is trained by feeding positive, negative and anchor sample into three separate \acp{DNN} $\mathcal C$ with weight sharing and using triplet loss with margin $M = 1$.

\subsection{Performance Metrics for Channel Charting}
Throughout this work, we will repeatedly evaluate channel charts on three performance metrics commonly used in channel charting literature \cite{studer_cc} \cite{siamese_cc}:
\begin{itemize}
    \item \textbf{\ac{CT}} and \textbf{\ac{TW}} \cite{trustworthiness_continuity} are two measures, normalized to the range $[0, 1]$, for the preservation of \emph{local} neighborhoods.
    Figuratively speaking, a high \ac{CT} indicates that many neighborhood relationships in physical space are preserved in the channel chart.
    A high \ac{TW} value, on the other hand, indicates that the channel chart does not contain many additional false neighborhood relationships, i.e., ones which are not present in physical space.
    We use \ac{CT} and \ac{TW} as defined in \cite{studer_cc} and adopt $K = 0.05 \cdot N$ as neighborhood size.
    \item \textbf{\ac{KS}}, first applied to channel charting in \cite{huang2019improving}, is a measure for the preservation of the global channel chart structure.
    It is also bounded to range $[0, 1]$, with $0$ indicating the best and $1$ indicating the worst possible performance.
    We adopt the scaling from \cite{triplet_cc}.
\end{itemize}

\section{Measured Datasets}
\label{sec:datasets}




\newcolumntype{M}[1]{>{\centering\arraybackslash}m{#1}}

\begin{figure*}
    \centering
    \begin{tabular}{| m{2em} | @{\hskip 0.8cm} M{5cm} M{5cm} M{5cm} |}
        \hline & & & \\[-0.8em]
        & \textbf{``Indoor'' Dataset} & \textbf{``Distributed'' Dataset} & \textbf{``Industrial'' Dataset} \\ & & & \\[-0.9em] \hline
        \rotatebox{90}{\parbox{2.5cm}{\centering \vspace{1ex}\leavevmode\newline Ground Truth}} & \begin{tikzpicture}[trim axis right, trim axis left]
            \begin{axis}[
                width=3.7cm,
                height=3.7cm,
                scale only axis,
                xmin=-6.5,
                xmax=0.5,
                ymin=-3.5,
                ymax=4,
                xlabel={Coordinate $x_1$ [m]},
                ylabel={Coordinate $x_2$ [m]}
            ]
                \addplot[thick,blue] graphics[xmin=-6.5,ymin=-3.5,xmax=0.5,ymax=4] {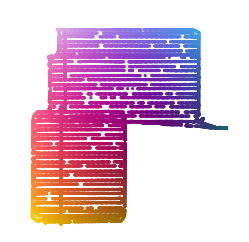};
            \end{axis}
        \end{tikzpicture} & \begin{tikzpicture}[trim axis right, trim axis left]
            \begin{axis}[
                width=3.7cm,
                height=3.7cm,
                scale only axis,
                xmin=-5.5,
                xmax=0.5,
                ymin=-1,
                ymax=6,
                xlabel={Coordinate $x_1$ [m]},
                ylabel={Coordinate $x_2$ [m]}
            ]
                \addplot[thick,blue] graphics[xmin=-5.5,ymin=-1,xmax=0.5,ymax=6] {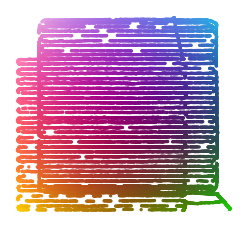};
            \end{axis}
        \end{tikzpicture} & \begin{tikzpicture}[trim axis right, trim axis left]
            \begin{axis}[
                width=3.7cm,
                height=3.7cm,
                scale only axis,
                xmin=-12,
                xmax=0,
                ymin=-2,
                ymax=11,
                xlabel={Coordinate $x_1$ [m]},
                ylabel={Coordinate $x_2$ [m]}
            ]
                \addplot[thick,blue] graphics[xmin=-12,ymin=-2,xmax=0,ymax=11] {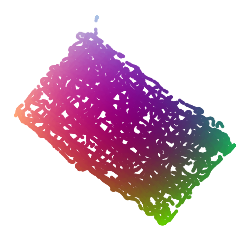};
            \end{axis}
        \end{tikzpicture} \\ \hline
        \rotatebox{90}{\parbox{2.5cm}{\centering State-of-the-Art\\Channel Chart}} &
        \begin{center}
            \footnotesize CT = 0.9486, TW = 0.8860, KS = 0.5268
        \end{center}
        \begin{tikzpicture}[trim axis right, trim axis left]
            \begin{axis}[
                width=3.7cm,
                height=3.7cm,
                scale only axis,
                xmin=-20,
                xmax=20,
                ymin=-18,
                ymax=18,
                xlabel={Latent Variable $z_1$},
                ylabel={Latent Variable $z_2$}
            ]
                \addplot[thick,blue] graphics[xmin=-20,ymin=-18,xmax=20,ymax=18] {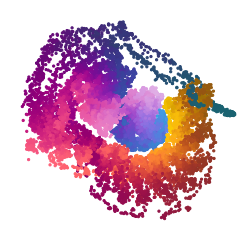};
            \end{axis}
        \end{tikzpicture} &
        \begin{center}
            \footnotesize CT = 0.9642, TW = 0.9295, KS = 0.3354
        \end{center}
        \begin{tikzpicture}[trim axis right, trim axis left]
            \begin{axis}[
                width=3.7cm,
                height=3.7cm,
                scale only axis,
                xmin=-17,
                xmax=17,
                ymin=-17,
                ymax=17,
                xlabel={Latent Variable $z_1$},
                ylabel={Latent Variable $z_2$}
            ]
        
                \addplot[thick,blue] graphics[xmin=-17,ymin=-17,xmax=17,ymax=17] {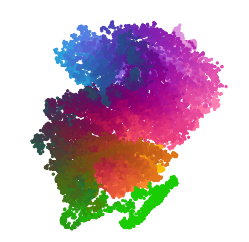};
            \end{axis}
        \end{tikzpicture} &
        \begin{center}
            \footnotesize CT = 0.9401, TW = 0.8790, KS = 0.3397
        \end{center}
        \begin{tikzpicture}[trim axis right, trim axis left]
            \begin{axis}[
                width=3.7cm,
                height=3.7cm,
                scale only axis,
                xmin=-15,
                xmax=15,
                ymin=-16.5,
                ymax=17,
                xlabel={Latent Variable $z_1$},
                ylabel={Latent Variable $z_2$}
            ]
                
                \addplot[thick,blue] graphics[xmin=-15,ymin=-16.5,xmax=15,ymax=17] {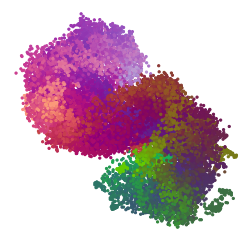};
            \end{axis}
        \end{tikzpicture} \\ \hline
        \rotatebox{90}{\parbox{2.5cm}{\centering Genie-aided\\Channel Chart}} &
        \begin{center}
            \footnotesize CT = 0.9935, TW = 0.9937, KS = 0.0880
        \end{center}
        \begin{tikzpicture}[trim axis right, trim axis left]
            \begin{axis}[
                width=3.7cm,
                height=3.7cm,
                scale only axis,
                xmin=-5.5,
                xmax=5,
                ymin=-6,
                ymax=6.5,
                xlabel={Latent Variable $z_1$},
                ylabel={Latent Variable $z_2$}
            ]
                
                \addplot[thick,blue] graphics[xmin=-5.5,ymin=-6,xmax=5,ymax=6.5] {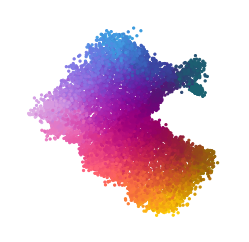};
            \end{axis}
        \end{tikzpicture} &
        \begin{center}
            \footnotesize CT = 0.9928, TW = 0.9930, KS = 0.0938
        \end{center}
        \begin{tikzpicture}[trim axis right, trim axis left]
            \begin{axis}[
                width=3.7cm,
                height=3.7cm,
                scale only axis,
                xmin=-6,
                xmax=6.5,
                ymin=-5,
                ymax=5.5,
                xlabel={Latent Variable $z_1$},
                ylabel={Latent Variable $z_2$}
            ]
                \addplot[thick,blue] graphics[xmin=-6,ymin=-5,xmax=6.5,ymax=5.5] {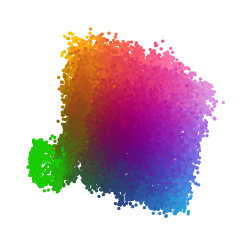};
            \end{axis}
        \end{tikzpicture} &
        \begin{center}
            \footnotesize CT = 0.9942, TW = 0.9939, KS = 0.0827
        \end{center}
        \begin{tikzpicture}[trim axis right, trim axis left]
            \begin{axis}[
                width=3.7cm,
                height=3.7cm,
                scale only axis,
                xmin=-7.5,
                xmax=7.5,
                ymin=-7,
                ymax=5.5,
                xlabel={Latent Variable $z_1$},
                ylabel={Latent Variable $z_2$}
            ]
                
                \addplot[thick,blue] graphics[xmin=-7.5,ymin=-7,xmax=7.5,ymax=5.5] {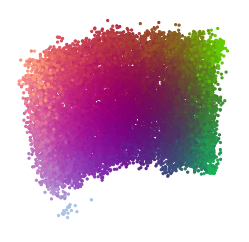};
            \end{axis}
        \end{tikzpicture} \\ \hline
        \rotatebox{90}{\parbox{4cm}{\centering Simulated Trajectories\\Channel Chart, $r = 30\,000$}} &
        \begin{center}
            \footnotesize CT = 0.9947, TW = 0.9943, KS = 0.0974
        \end{center}
        \begin{tikzpicture}[trim axis right, trim axis left]
            \begin{axis}[
                width=3.7cm,
                height=3.7cm,
                scale only axis,
                xmin=-7,
                xmax=8,
                ymin=-6.5,
                ymax=5,
                xlabel={Latent Variable $z_1$},
                ylabel={Latent Variable $z_2$}
            ]
                \addplot[thick,blue] graphics[xmin=-7,ymin=-6.5,xmax=8,ymax=5] {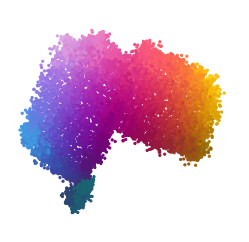};
            \end{axis}
        \end{tikzpicture} &
        \begin{center}
            \footnotesize CT = 0.9941, TW = 0.9942, KS = 0.0783
        \end{center}
        \begin{tikzpicture}[trim axis right, trim axis left]
            \begin{axis}[
                width=3.7cm,
                height=3.7cm,
                scale only axis,
                xmin=-6.5,
                xmax=7,
                ymin=-6.5,
                ymax=6,
                xlabel={Latent Variable $z_1$},
                ylabel={Latent Variable $z_2$}
            ]
                \addplot[thick,blue] graphics[xmin=-6.5,ymin=-6.5,xmax=7,ymax=6] {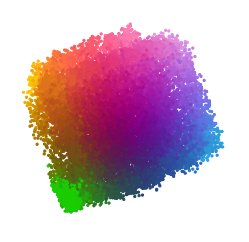};
            \end{axis}
        \end{tikzpicture} &
        \begin{center}
            \footnotesize CT = 0.9636, TW = 0.9384, KS = 0.2605
        \end{center}
        \begin{tikzpicture}[trim axis right, trim axis left]
            \begin{axis}[
                width=3.7cm,
                height=3.7cm,
                scale only axis,
                xmin=-8,
                xmax=6.5,
                ymin=-7,
                ymax=7,
                xlabel={Latent Variable $z_1$},
                ylabel={Latent Variable $z_2$}
            ]
                \addplot[thick,blue] graphics[xmin=-8,ymin=-7,xmax=6.5,ymax=7] {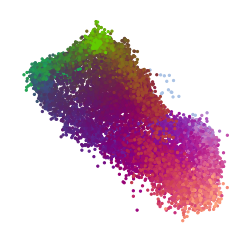};
            \end{axis}
        \end{tikzpicture} \\ \hline
    \end{tabular}
    \caption{Colored ground truth positions and different channel charts (with colorings preserved) including performance metrics generated from \ac{CSI}: Computed with ``state-of-the-art'' method, genie-aided triplet generation and from simulated trajectories.}
    \label{fig:charts}
\end{figure*}

We analyze three datasets captured by our channel sounder called \ac{DICHASUS}, with a carrier frequency of $1.272\,\mathrm{GHz}$, $W = 1024$ \ac{OFDM} subcarriers and a bandwidth of $50\,\mathrm{MHz}$:
\begin{itemize}
    \item \textbf{Indoor}: Subset of \emph{dichasus-015x} \cite{dataset-dichasus-015x} with $N = 13496$ datapoints, measured in an indoor office environment with a single $B = 32$-antenna uniform planar array. The transmitter moves along a meandering trajectory.
    \item \textbf{Distributed}: Subset of \emph{dichasus-0d5x} \cite{dataset-dichasus-0d5x} with $N = 22332$ datapoints, measured in the same indoor environment as the ``Indoor'' dataset, but with two separate 16-antenna uniform planar arrays ($B = 32$ in total). The transmitter moves along a meandering trajectory.
    \item \textbf{Industrial}: Subset of \emph{dichasus-cb0x} \cite{dataset-dichasus-cb0x} with $N = 19919$ datapoints, measured in a factory building with a single antenna array, with data from $B = 21$ antennas in the dataset. The transmitter moves along different meandering and pseudorandom trajectories with high speed.
\end{itemize}

To reduce computational cost and in contrast to Section \ref{sec:featureengineering}, only a mean value, computed over the channel coefficients of 8 subcarriers in the center of the band, is used for $\mathbf h_n$:
\[
    \left[ \mathbf h_n \right]_b = \frac{1}{8} \sum_{w = 508}^{515} \left[\mathbf S_{n}\right]_{b, w},
\]
where $\left[\mathbf S_{n}\right]_{b, w}$ denotes the channel coefficient measured by antenna $b$ on subcarrier $w$.
We continue with Feature Engineering as explained in Section \ref{sec:featureengineering}, mapping $\mathbf h_n \in \mathbb{C}^{B}$ to the feature vector $\mathbf f_n \in \mathbb C^{B^2}$ that is provided to the \ac{DNN} for training and evaluation, as explained in Section \ref{sec:stateoftheart}.
The ground truth position labels and resulting channel charts for all datasets are shown in Fig. \ref{fig:charts} in the first and second rows, respectively.
All charts, judging by performance metrics and appearance, are acceptable, but fail to learn the global geometry.
For example, the rectangular shapes in the ground truth are not recognizable in the channel charts.
In the following, we will show that, without changing the training procedure, performance could be significantly improved through better triplet selection.

\section{Triplet Generation}
\label{sec:tripletgeneration}

Triplet selection in state-of-the-art channel charting is time-based and therefore dependent on the particular trajectory taken by the \ac{UE}.
To understand if different trajectories would have yielded better channel charts with the same \ac{DNN} training technique, we can leverage the ground truth position labels contained in \ac{DICHASUS} datasets.
We compare an idealized genie-aided triplet selection technique to a method based on large numbers of simulated (but realistic) trajectories.

\subsection{Distance-Based Triplet Selection (Genie-Aided)}
\begin{figure}[H]
    \centering
    \begin{tikzpicture}

\begin{axis}[
tick align=outside,
tick pos=left,
grid=major,
grid style={dashed},
xlabel={Max. positive sample distance $d_\mathrm{c}$ [m]},
xmajorgrids,
xmin=0.375, xmax=3.125,
xtick style={color=black},
ylabel={CT \& TW},
ymin=0.93, ymax=1.01,
ytick={0.94, 0.97, 1},
ytick style={color=black},
height = 4.5cm,
width = 7cm
]
\addplot [semithick, mittelblau, mark=triangle, mark size = 2.5, mark options={opacity = 0.8}]
table {%
0.5 0.993600010871887
0.800000011920929 0.994899988174438
1 0.998199999332428
1.5 0.997699975967407
2 0.996699988842011
2.5 0.992999970912933
3 0.981899976730346
};
\label{ct}
\addplot [semithick, apfelgruen, mark=square, mark size = 2.5, mark options={opacity = 0.8}]
table {%
0.5 0.949100017547607
0.800000011920929 0.980400025844574
1 0.998199999332428
1.5 0.997699975967407
2 0.996699988842011
2.5 0.99260002374649
3 0.980300009250641
};
\label{tw}
\end{axis}

\begin{axis}[
axis y line=right,
tick align=outside,
xmin=0.375, xmax=3.125,
xmajorticks=false,
ylabel={KS},
ymin=0.0, ymax=0.4,
ytick={0.05, 0.20, 0.35},
yticklabel style={
        /pgf/number format/fixed,
        /pgf/number format/precision=2
},
scaled y ticks=false,
ytick pos=right,
ytick style={color=black},
yticklabel style={anchor=west},
height = 4.5cm,
width = 7cm,
legend cell align={left},
axis line style = {-},
legend style={
  fill opacity=0.8,
  draw opacity=1,
  text opacity=1,
  at={(0.5,1.03)},
  anchor=south,
  draw=white!80!black,
  column sep=5pt
},
legend columns = 3
]
\addplot [semithick, orange, mark=*, mark size = 2, mark options={opacity = 0.8}]
table {%
0.5 0.361299991607666
0.800000011920929 0.269699990749359
1 0.137600004673004
1.5 0.0537000000476837
2 0.0559000000357628
2.5 0.0798999965190887
3 0.128800004720688
};
\addlegendentry{KS}

\addlegendimage{/pgfplots/refstyle=ct}\addlegendentry{CT}
\addlegendimage{/pgfplots/refstyle=tw}\addlegendentry{TW}

\end{axis}

\draw ({$(current bounding box.south west)!0.5!(current bounding box.south east)$}|-{$(current bounding box.south west)!0.98!(current bounding box.north west)$}) node[
  scale=0.6,
  anchor=north,
  text=black,
  rotate=0.0
] {};
\end{tikzpicture}
    \caption{Performance metrics as a function of $d_\mathrm{c}$, genie-aided triplet selection evaluated on ``Indoor'' dataset}
    \label{fig:genie_indoor_sweep_dc}
\end{figure}
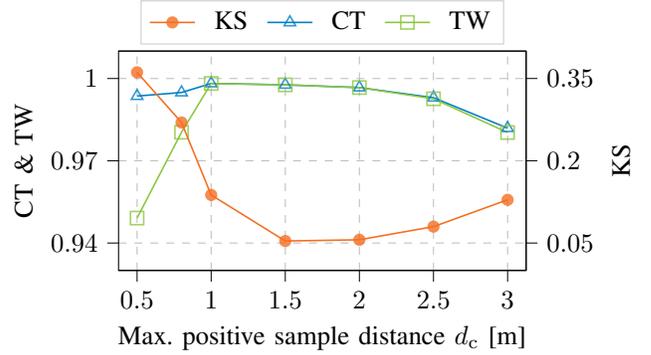

Triplets are selected by randomly drawing an anchor point $(\mathbf S_\mathrm{anchor}, \mathbf x_\mathrm{anchor}, t_\mathrm{anchor})$ and negative sample $(\mathbf S_\mathrm{neg}, \mathbf x_\mathrm{neg}, t_\mathrm{neg})$ from the whole dataset and a positive sample from the set of all datapoints $(\mathbf S_n, \mathbf x_n, t_n)$ for which $\lVert \mathbf x_n - \mathbf x_\mathrm{anchor} \rVert \leq d_\mathrm{c}$, where $d_\mathrm{c}$ is the maximum spatial distance between anchor point and positive sample.
All random selections are made with uniform probability.
Experiments have shown that imposing $\lVert \mathbf x_\mathrm{anchor} - x_\mathrm{neg} \rVert > d_c$ as an additional restriction only marginally improves performance.
It was found that $d_c = 1.5\,\mathrm{m}$ is a good choice for the ``Indoor'' dataset with regards to the performance metrics, as illustrated in Fig. \ref{fig:genie_indoor_sweep_dc}.
The resulting channel charts, shown in the third row of Fig. \ref{fig:charts}, demonstrate that local and global geometry is preserved when training on genie-generated triplets.

\subsection{Time-Based Triplet Selection on Simulated Trajectories}
\begin{figure*}
    \centering
    \begin{subfigure}[b]{0.26\textwidth}
        \centering
        \scalebox{0.9}{
            \begin{tikzpicture}
    \begin{axis}[
        width=3.7cm,
        height=3.7cm,
        scale only axis,
        xmin=-6,
        xmax=0.5,
        ymin=-3.5,
        ymax=3.5,
        xlabel={$x$ coordinate [m]},
        ylabel={$y$ coordinate [m]}
        ]
        \addplot[thick,blue] graphics[xmin=-6,ymin=-3.5,xmax=0.5,ymax=3.5] {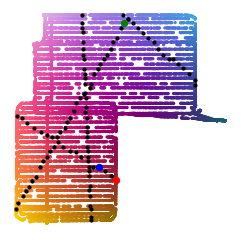};
    \end{axis}
\end{tikzpicture}
        }
        \vspace{-0.4cm}
        \caption{Conceptual illustration of simulated trajectories (``Indoor'' dataset)}
        \label{fig:sim_trajectories_example}
    \end{subfigure}
    \hspace{0.01\textwidth}
    \begin{subfigure}[b]{0.26\textwidth}
        \centering
        {\hspace{0.42cm} \scriptsize CT = 0.9855, TW = 0.9839, KS = 0.1193} \newline \\[-0.5em]
        \scalebox{0.9}{
            \begin{tikzpicture}
    \begin{axis}[
        width=3.7cm,
        height=3.7cm,
        scale only axis,
        xmin=-6,
        xmax=5.5,
        ymin=-5,
        ymax=4.5,
        xlabel={Latent Variable $z_1$},
        ylabel={Latent Variable $z_2$}
        ]
        \addplot[thick,blue] graphics[xmin=-6,ymin=-5,xmax=5.5,ymax=4.5] {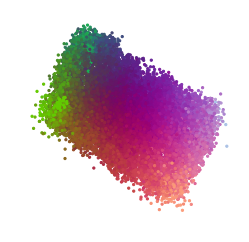};
    \end{axis}
\end{tikzpicture}
        }
        \vspace{-0.4cm}
        \caption{Simulated trajectories channel chart, ``Industrial'' dataset, $T_c = 3$}
        \label{fig:sim_trajectories_industrial_larger_tc}
    \end{subfigure}
    \hspace{0.01\textwidth}
    \begin{subfigure}[b]{0.42\textwidth}
        \centering
        \scalebox{0.9}{
            \begin{tikzpicture}

\begin{axis}[
log basis x={10},
tick align=outside,
tick pos=left,
grid=major,
grid style={dashed},
xlabel={number of simulated trajectories $r$},
xmajorgrids,
xmin=6.70106664593439, xmax=44768.9921397833,
xmode=log,
xtick style={color=black},
ylabel={CT \& TW},
ymin=0.65, ymax=1.05,
ytick style={color=black},
width = 7cm,
height = 5cm
]
\addplot [semithick, mittelblau, mark=triangle, mark size=2.5, mark options={opacity = 0.8}]
table {%
10 0.717299997806549
30 0.797299981117249
100 0.914099991321564
300 0.968500018119812
1000 0.989600002765656
2000 0.987399995326996
5000 0.994199991226196
10000 0.994799971580505
30000 0.995000004768371
};
\label{ctb}
\addplot [semithick, apfelgruen, mark=square, mark size=2.5, mark options={opacity = 0.8}]
table {%
10 0.675100028514862
30 0.783399999141693
100 0.913900017738342
300 0.952099978923798
1000 0.987999975681305
2000 0.980199992656708
5000 0.994000017642975
10000 0.994400024414062
30000 0.994799971580505
};
\label{twb}
\end{axis}

\begin{axis}[
legend style={
  fill opacity=0.8,
  draw opacity=1,
  text opacity=1,
  at={(0.5,1.03)},
  anchor=south,
  draw=white!80!black,
  column sep=5pt
},
legend columns = 3,
axis y line=right,
log basis x={10},
tick align=outside,
xmin=6.70106664593439, xmax=44768.9921397833,
xmode=log,
xmajorticks=false,
xtick style={draw=none},
ylabel={KS},
axis line style = {-},
ymin=0.0, ymax=0.8,
ytick={0.1, 0.3, 0.5, 0.7},
ytick pos=right,
ytick style={color=black},
yticklabel style={anchor=west},
width = 7cm,
height = 5cm
]
\addplot [semithick, orange, mark=*, mark size=2.5, mark options={opacity = 0.8}]
table {%
10 0.571099996566772
30 0.456900000572204
100 0.30349999666214
300 0.240099996328354
1000 0.142000004649162
2000 0.195600003004074
5000 0.09009999781847
10000 0.0864000022411346
30000 0.0834999978542328
};
\addlegendentry{KS}

\addlegendimage{/pgfplots/refstyle=ctb}\addlegendentry{CT}
\addlegendimage{/pgfplots/refstyle=twb}\addlegendentry{TW}
\end{axis}

\end{tikzpicture}
        }
        \vspace{-0.4cm}
        \caption{Performance as a function of the number of simulated trajectories, evaluated on ``Indoor'' dataset}
        \label{fig:numberoftrajectories}
    \end{subfigure}
    \vspace{-0.1cm}
    \caption{Several illustrations pertaining the use of simulated trajectories for triplet generation}
    \label{fig:fake_trajectories_results}
    \vspace{-0.2cm}
\end{figure*}
\vspace{-0.2cm}

A set of $r$ simulated, but realistic \ac{UE} trajectories with constant velocity $v = 1 \, \sfrac{\mathrm m}{\mathrm s}$ through the dataset is generated and triplets are drawn from this set according to the state-of-the-art time-based triplet selection with $T_\mathrm{c} = 1.5\,\mathrm{s}$.
This is best explained with Fig. \ref{fig:sim_trajectories_example}, where four exemplary simulated trajectories were generated (black), anchor (red) and positive sample (blue) were randomly drawn within $T_\mathrm{c}$ on the same trajectory, and the negative sample (green) is chosen from a random trajectory.
We only consider disjoint straight lines as potential trajectories, experiments with other trajectory shapes such as B\'ezier curves did not yield any performance improvements.
Compared to triplet selection based on the \ac{UE}'s real trajectory, using simulated trajectories makes it possible to remove biases introduced during data acquisition (e.g., meandering \ac{UE} paths) and to better understand desirable properties of trajectories.
For large numbers of simulated trajectories ($r \to \infty$), time-based selection on simulated trajectories becomes similar (but never identical) to genie-aided triplet selection.


With $r = 30\,000$, performance comparable to genie-aided triplet selection is achievable as shown in the fourth row of Fig. \ref{fig:charts}, except for the ``Industrial'' dataset.
Datapoints in that dataset are spread over a larger area, thus setting $T_\mathrm{c} = 3$ improves the result significantly, as illustrated in Fig. \ref{fig:sim_trajectories_industrial_larger_tc}.
For the other datasets, the results are almost as good as those produced with genie-aided triplet selection.
Fig. \ref{fig:numberoftrajectories}, which was generated based on the ``Indoor'' dataset, shows that channel charting performance is highly dependent on the number of simulated trajectories.
We conclude that a dataset which contains a large variety of paths is essential for channel charting.
\section{Transferring Channel Charts}
\label{sec:transfer}
\begin{figure}
    \centering
    \begin{subfigure}[t]{0.49\columnwidth}
        \centering
        {\scriptsize CT = 0.9743, TW = 0.9716, KS = 0.1551} \newline \\[-0.3em]
        \scalebox{0.9}{
            \begin{tikzpicture}[trim axis right, trim axis left]
    \begin{axis}[
        width=3.7cm,
        height=3.7cm,
        scale only axis,
        xmin=-5.5,
        xmax=5,
        ymin=-6,
        ymax=6.5,
        xlabel = {Latent Variable $z_1$},
        ylabel = {Latent Variable $z_2$},
        ylabel shift = -8 pt,
        xlabel shift = -4 pt
    ]
        \addplot[thick,blue] graphics[xmin=-5.5,ymin=-6,xmax=5,ymax=6.5] {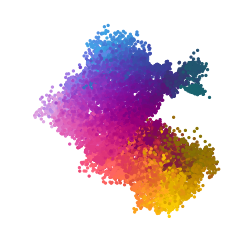};
    \end{axis}
\end{tikzpicture}
        }
        \vspace{-0.1cm}
        \caption{Same environment}
        \label{fig:genie_other_dataset}
    \end{subfigure}
    \begin{subfigure}[t]{0.49\columnwidth}
        \centering
        {\scriptsize CT = 0.8590, TW = 0.8116, KS = 0.3843} \newline \\[-0.3em]
        \scalebox{0.9}{
            \begin{tikzpicture}[trim axis right, trim axis left]
    \begin{axis}[
        width=3.7cm,
        height=3.7cm,
        scale only axis,
        xmin=-4.5,
        xmax=4.5,
        ymin=-5,
        ymax=5.5,
        xlabel = {Latent Variable $z_1$},
        ylabel = {Latent Variable $z_2$},
        ylabel shift = -8 pt,
        xlabel shift = -4 pt
    ]
        \addplot[thick,blue] graphics[xmin=-4.5,ymin=-5,xmax=4.5,ymax=5.5] {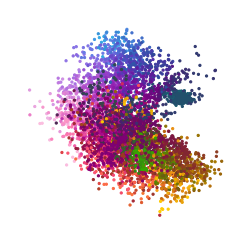};
    \end{axis}
\end{tikzpicture}
        }
        \vspace{-0.1cm}
        \caption{Modified environment}
        \label{fig:genie_other_environment}
    \end{subfigure}
    \vspace{-0.1cm}
    \caption{Transferred channel charts: $\mathcal C$ is evaluated on different datasets after training on the ``Indoor'' dataset}
\end{figure}

We apply the forward charting function $\mathcal C$ previously learned for the ``Indoor'' dataset after genie-aided triplet selection (Fig. \ref{fig:charts}, third row, first column) to different, previously unseen data.
Fig. \ref{fig:genie_other_dataset} shows the channel chart obtained by applying $\mathcal C$ to a \ac{CSI} dataset captured in the exact same radio environment at a later point in time.
Despite having never seen any datapoint from this set, $\mathcal C$ can still extract the local and global geometry.
In another experiment, $\mathcal C$ is evaluated on a different dataset in a similar environment, but with an additional obstacle in the line-of-sight path \cite{dataset-dichasus-005x}.
As illustrated in Fig. \ref{fig:genie_other_environment}, $\mathcal C$ still recognizes contiguous regions, but performance is degraded.
\section{Summary and Outlook}
We have successfully applied triplet-based channel charting to multiple datasets acquired by a massive \ac{MIMO} channel sounder.
We investigated the importance of triplet selection and showed the possibility of transferring learned forward charting functions to unseen data.
The transferability of channel charts to new data motivates a potential pre-training of the \ac{DNN} that implements the forward charting function on model-generated data, as in \cite{Arnold2018OnDL}.
With this paper, we highlight the importance of large and varied training datasets and provide a framework for evaluating neural network training techniques independent of the true \ac{UE} trajectory.

\vspace{-0.1cm}

\bibliographystyle{IEEEtran}
\bibliography{IEEEabrv,references}

\end{document}